\documentclass[12pt,a4paper]{conference}

\usepackage{fancyhdr}
\usepackage{graphicx,amsmath,amssymb,cite}
\usepackage{multind}
\usepackage{epsfig}
\makeindex{author} \makeindex{subject}

\newcommand{\psp}{\psi(2S)}
\newcommand{\llp}{\llb\pi^0}
\newcommand{\lle}{\llb\eta}
\newcommand{\jpsi}{J/\psi}
\newcommand{\lm}{\Lambda}
\newcommand{\bcl}{\begin{center}}
\newcommand{\llb}{\Lambda\bar{\Lambda}}
\newcommand{\GG}{\gamma\gamma}
\newcommand{\ar}{\rightarrow}
\newcommand{\splb}{\Sigma^+\pi^-\bar{\Lambda}}
\newcommand{\sbpl}{\bar{\Sigma}^-\pi^+\lm}
\newcommand{\ecl}{\end{center}}
\newcommand{\lmb}{\bar{\Lambda}}

\newcommand{\psip}{\psi(2S)}

\newcommand{\jpsito}{J/\psi\rightarrow}

\newcommand{\beq}{\begin{equation}}
\newcommand{\eeq}[1]{\label{#1}\end{equation}}
\newcommand{\eeqn}{\end{equation}}

\newcommand{\beqa}{\begin{eqnarray}}
\newcommand{\eeqa}[1]{\label{#1}\end{eqnarray}}
\newcommand{\eeqan}{\end{eqnarray}}

\let\bar=\overbar

\newcommand{\Dslash}{\not{\hbox{\kern-4pt $D$}}}
\newcommand{\dslash}{\not{\hbox{\kern-2pt $\del$}}}

\newcommand{\msb}{{\bar{\ssstyle M \kern -1pt S}}}

\begin{document}

\Chapter{Recent BES results and the BESIII upgrade}
           {Recent BES Results and the BESIII Upgrade}{F. A. Harris}
\vspace{-6 cm}\includegraphics[width=6 cm]{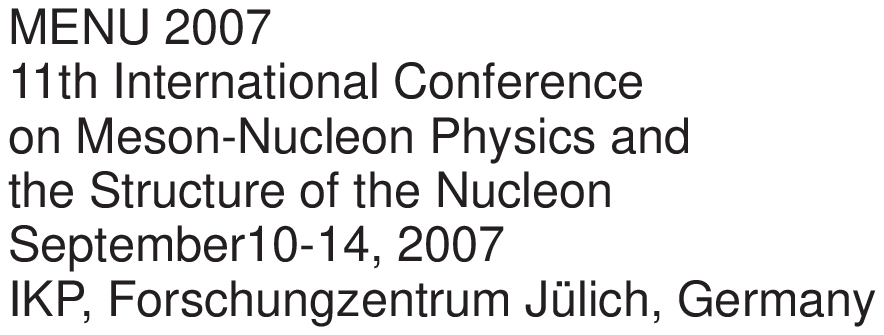}
\vspace{4 cm}

\addcontentsline{toc}{chapter}{{\it F. A. Harris}} \label{authorStart}

\begin{raggedright}

{\it F. A. Harris \\
for the BES Collaboration}\index{author}{Harris, F. A.}\\
Department of Physics and Astronomy\\
University of Hawaii\\
Honolulu, HI 96822\\
USA
\bigskip\bigskip

\end{raggedright}

\begin{center}
\textbf{Abstract}
\end{center}


Using 58 million $\jpsi$ and 14 million $\psi(2S)$ events collected by
the BESII detector at the BEPC, branching fractions or upper limits
for the decays $\jpsi$ and $\psp\ar\llp$ and $\lle$ are measured, and
the decays of $\jpsi$ and $\psip$ to ${n}{K^0_S}\bar{\Lambda}+c.c.$
are observed and measured for the first time.  Finally, $R$
measurement data taken with the BESII detector at center-of-mass
energies between 3.7 and 5.0 GeV are fitted to determine resonance
parameters of the high mass charmonium states, $\psi(3770)$,
$\psi(4040)$, $\psi(4160)$, and $\psi(4415)$.

The Beijing Electron Collider is being upgraded to a two-ring collider
(BEPCII) with a design luminosity of $1 \times 10^{33}$cm$^{-2}$
s$^{-1}$ at 3.89 GeV and will operate between 2 and 4.2 GeV in the
center of mass.  With this luminosity, the new BESIII detector will be
able to collect, for example, 10 billion $J/\psi$ events in one year
of running.  BEPCII and BESIII are currently nearing completion, and
commissioning of both is expected to begin in mid-2008.

\section{Introduction}
In this paper, some
recent BESII results are reported based on 58 million $\jpsi$ and 14
million $\psi(2S)$ events collected by the BESII detector at the BEPC,
and the status of BESIII/BEPCII is summarized.  For much more detail,
see the references.

\section{Recent Results}

\subsection{\boldmath $\jpsi$ and $\psp$ decays into $\llp$ and $\lle$}

The isospin violating decay $\jpsi\ar\llp$ was studied by
DM2~\cite{np3} and BESI~\cite{np4}, and its average branching fraction
is ${\cal B}(\jpsi\ar\llp)=(2.2\pm 0.6)\times
10^{-4}$~\cite{pdg06}. However, the isospin conserving process
$\jpsi\ar\lle$ has not been reported, and there are no measurements
for $\llp$ and $\lle$ decays of $\psp$.

Here $\jpsi\ar\llp$, $\jpsi\ar\lle$,
$\psp\ar\llp$, and $\psp\ar\lle$, where $\lm$ decays to $\pi^- p$ and
$\pi^0$ and $\eta$ to $\GG$, are studied. Candidate events must have four good
charged tracks and at least two photons, two protons identified using
particle identification, a satisfactory four constraint kinematic fit,
and a $\pi p$ mass consistent with the $\Lambda$ mass. For
$\jpsi\ar\llp$, the decay lengths of $\lm$ and $\lmb$ in the $x-y$
plane must be larger than 0.05 m.

Possible backgrounds are studied using MC simulation.  The most
serious one is from $\jpsi\ar\Sigma^0\pi^0\lmb+c.c.$, which contains
$\llp$ with an additional photon in the final state.  The branching
fraction for this decay has not been previously measured. Since direct
measurement of $\jpsi\ar\Sigma^0\pi^0\lmb+c.c.$ is difficult, we
measure the branching fractions of its isospin partner and estimate
the branching fraction assuming isospin symmetry.

\begin{figure}[!htb]
\begin{center}
\begin{minipage}[t]{2.6in}
\centerline{\psfig{file=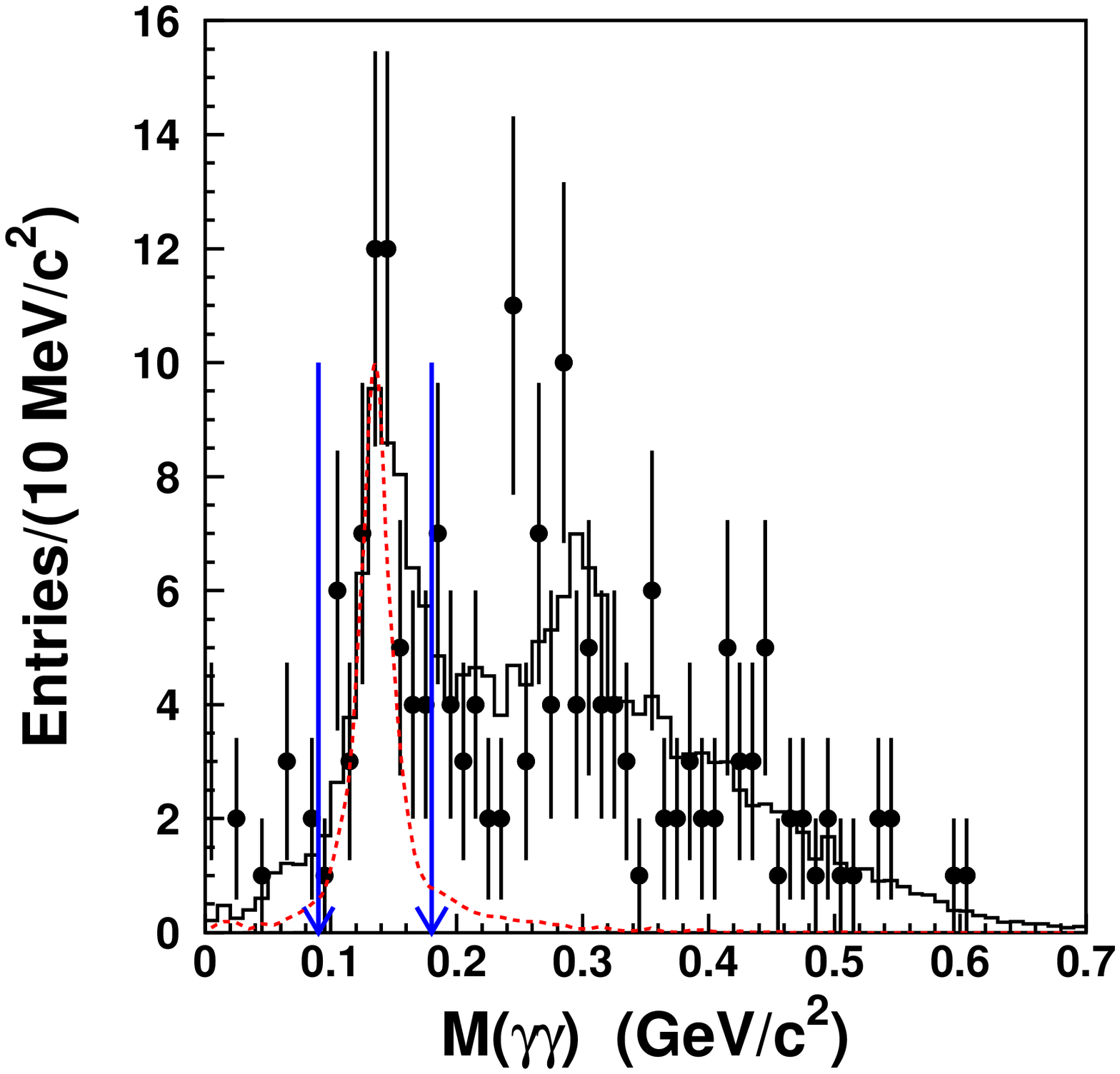,width=6cm,height=4.5cm}}
\caption{Invariant mass distribution of $M(\GG)$ for $\jpsi\ar\Lambda
  \bar{\Lambda} \pi^0 \ar\llb\GG$
  candidates (dots with error bars) and normalized backgrounds (solid
  histogram). The dashed curves shows the $\pi^0$ signal from MC
  simulated $\jpsi\ar\llp$. }
\label{nnfig5}
\end{minipage} \ \
\begin{minipage}[t]{2.60in}
\centerline{\psfig{file=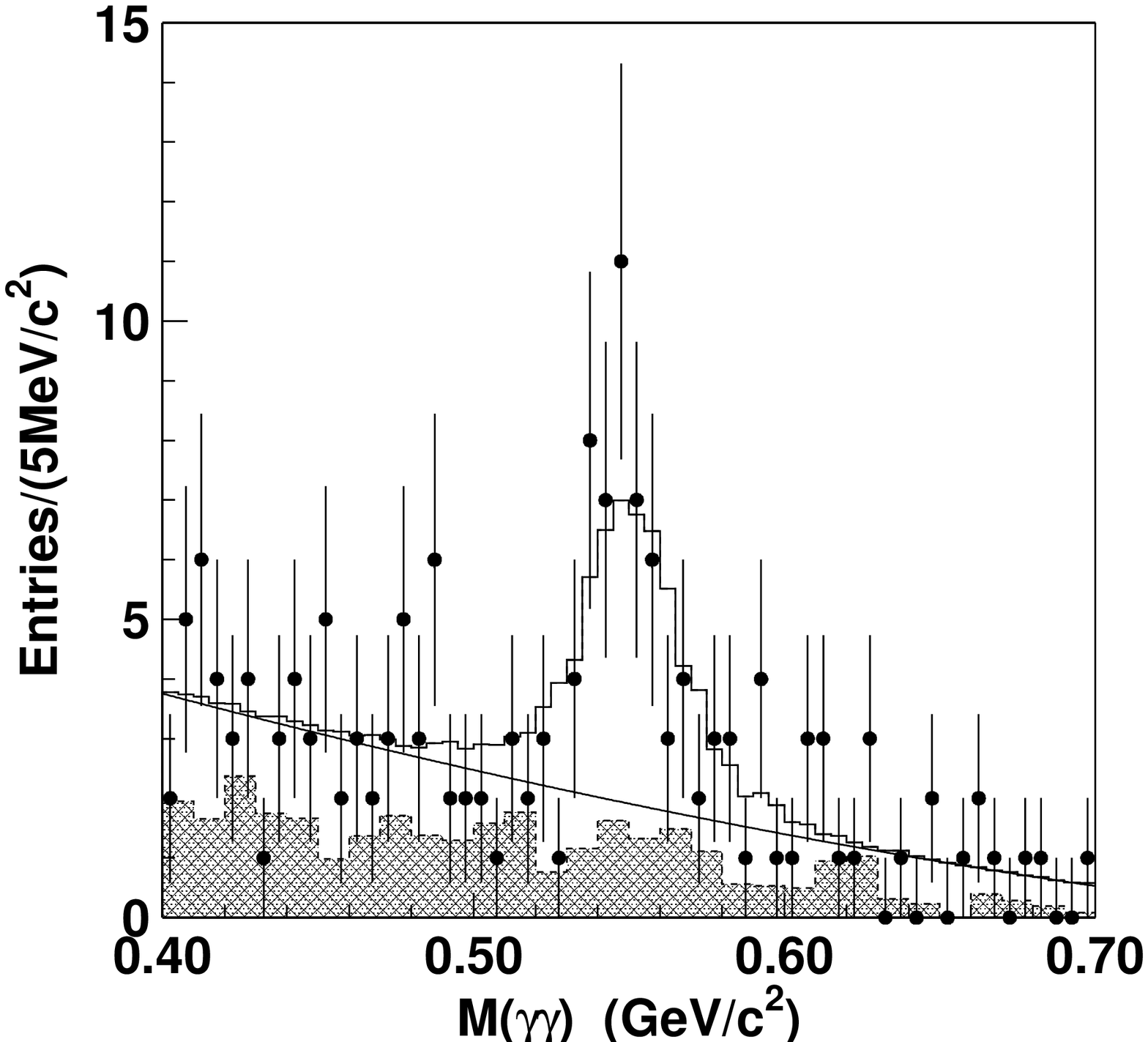,width=6cm,height=4.5cm}}
\caption{Fit to the $\GG$ invariant mass distribution of
$\jpsi\ar\Lambda \bar{\Lambda}\eta\ar\llb\GG$ candidate events.
Dots
with error bars are data, the hatched histogram is the normalized
background, and the
solid histogram is the fit to data using a histogram of the signal
shape from MC simulation plus a second order polynomial for
background.}
\label{nnfig9}
\end{minipage}
\end{center}
\end{figure}

\begin{figure}[!htb]
\begin{center}
\begin{minipage}[t]{2.6in}
\centerline{\psfig{file=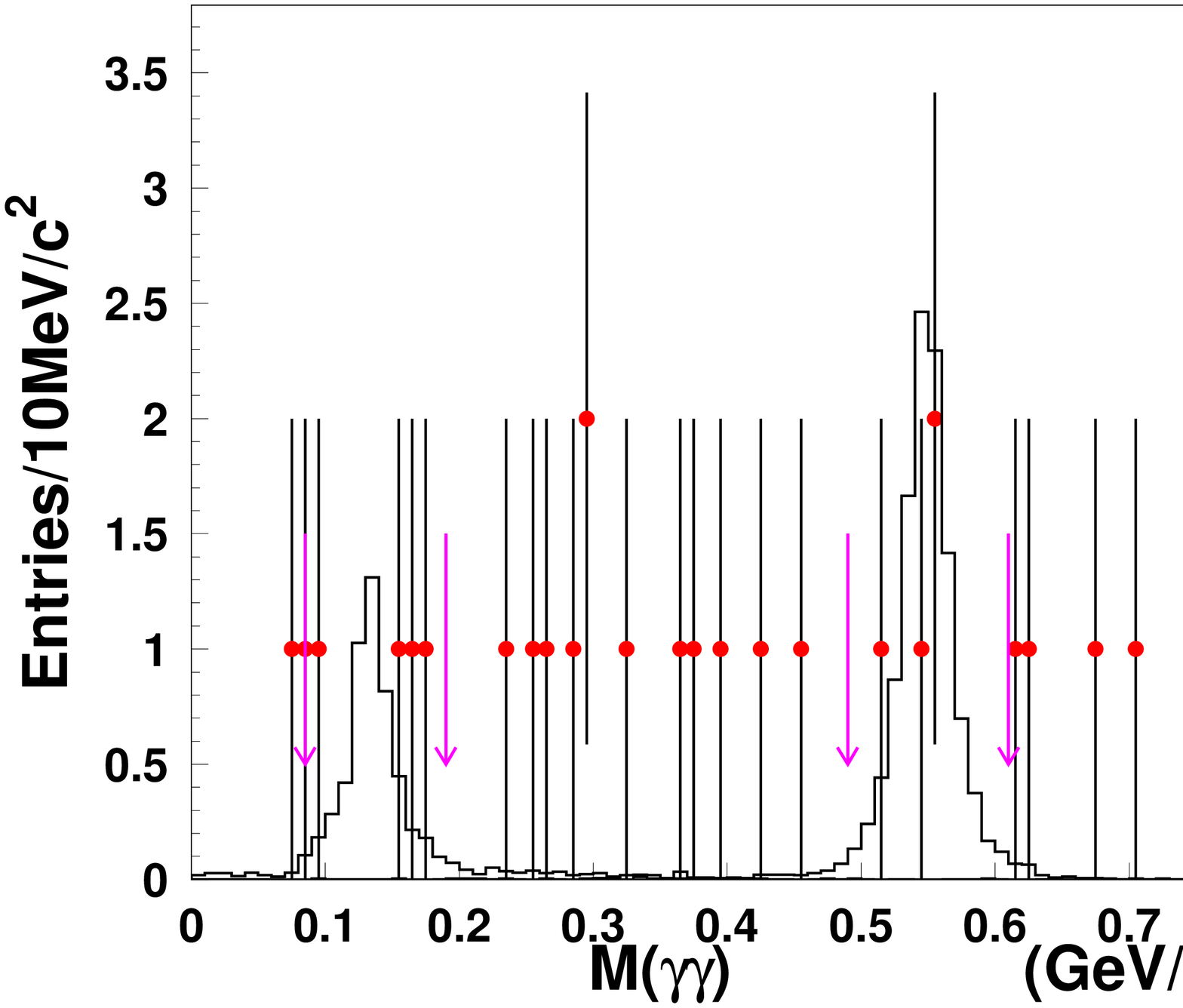,width=5.8cm,height=4.5cm}}
\caption{The $\GG$ invariant mass distribution for candidate
$\psp\ar\GG\llb$ events. Dots with error bars are data, and the
histograms are MC simulated signal events. The arrows indicate the
$\pi^0$ and $\eta$ signal regions.}
\label{nnfig13}
\end{minipage} \ \
\begin{minipage}[t]{2.60in}
\centerline{\hbox{\psfig{file=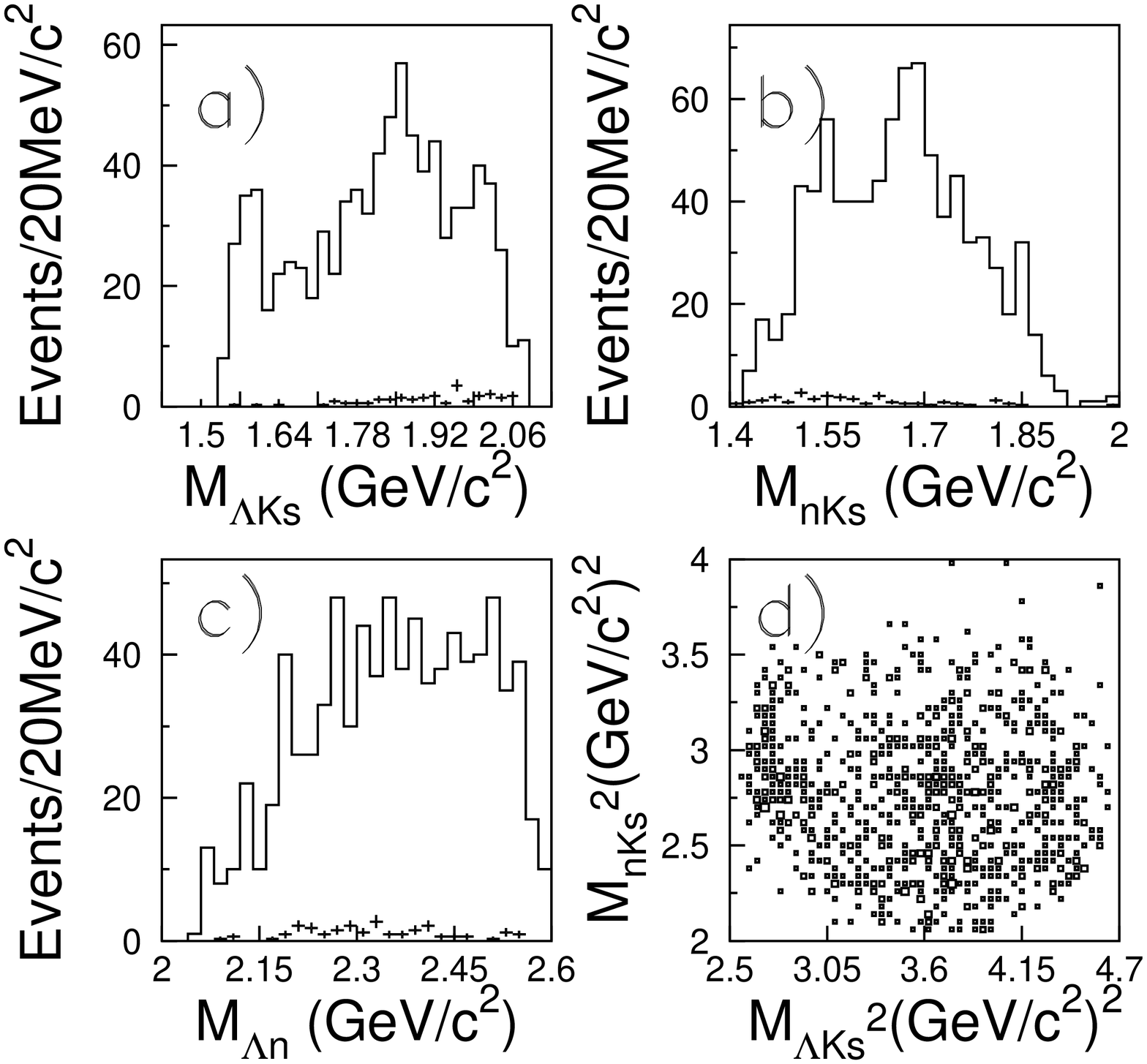,width=7cm,height=7cm}}}
\caption{The invariant mass spectra of (a) ${\Lambda K_S^0}$, (b) ${n K_S^0}$,
  and (c) ${\bar {\Lambda} n (\Lambda \bar n)}$, as well as (d)
  the Dalitz plot for candidate $J/\psi \to nK^0_S \Lambda
  + c.c.$ events after all selection criteria.
  The crosses show the sideband backgrounds.}
\label{Nstar}
\end{minipage}
\end{center}
\end{figure}

The histogram in Fig.~\ref{nnfig5} shows normalized backgrounds from
all background channels, and the dashed line in the figure shows the
$\pi^0$ signal from MC simulated $\jpsi\ar\llp$. The data in
Fig.~\ref{nnfig5} are consistent with background, and the upper limit
on the number of $\pi^0$ events from $\jpsi\ar\llp$ is determined.
 Figure~\ref{nnfig9} shows the the $\gamma \gamma$ invariant mass
distribution for $\jpsi\ar\Lambda \bar{\Lambda} \eta$ candidates, where
a clear $\eta$ signal is observed.  Figure~\ref{nnfig13} shows the
$\GG$ invariant mass distribution for $\psp\ar\llp$ and $\lle$ using
similar selection criteria, and no significant $\pi^0$ or $\eta$
signals are seen.

\begin{table*}
\caption{Measured branching fractions or upper limits at the 90\%
confidence level (C.L.).} 
\bcl
\footnotesize
\begin{tabular}{l|c|c|c}\hline\hline
Channels&Number of events&MC efficiency (\%)&Branching fraction ($\times 10^{-4}$)\\ \hline
$\jpsi\ar\llp$&$<11.2$&0.75&$<0.64$\\ 
$\jpsi\ar\lle$&$44\pm 10$&$1.8$&$2.62\pm 0.60\pm 0.44$\\ \hline
$\psp\ar\llp$&$<7.0$&2.5&$<0.49$\\ 
$\psp\ar\lle$&$<7.6$&2.9&$<1.2$\\ \hline
$\jpsi\ar\splb$&$335 \pm 22$&2.3&$7.70\pm 0.51\pm 0.83$ \\
$\jpsi\ar\sbpl$&$254 \pm 19$&1.8&$7.47\pm 0.56\pm 0.76$ \\ \hline
\end{tabular}
\label{branresult}
\ecl
\end{table*}

Table ~\ref{branresult} lists the results for $\jpsi$ and $\psp$ decay
into $\llp$ and $\lle$, as well as $\jpsi\ar\splb+c.c.$.  Except for
$\jpsi\ar\llp$ and $\jpsi\ar\splb+c.c.$, the results are first
measurements. Interestingly, the result of $\jpsi\ar\llp$
presented here is much smaller than those of DM2 and
BESI~\cite{np3,np4}. Previously, the large contaminations
from $\jpsi\ar\Sigma^0\pi^0\lmb+c.c.$ and
$\jpsi\ar\Sigma^+\pi^-\lmb+c.c.$ were not considered, resulting in a
large $\jpsi\ar\llp$ branching fraction. The small
branching fraction of $\jpsi\ar\llp$ and relatively large branching
fraction of $\jpsi\ar\lle$ measured here indicate that the isospin
violating decay in $\jpsi$ decays is suppressed while the isospin
conserving decays is favored, which is consistent with
expectation. For more detail, see Ref.~\cite{LLpi}.

\subsection{\boldmath $J/\psi$ and $\psi(2S)$ decaying to $nK^0_S \Lambda
  + c.c.$}

In 2004, BESII reported the observation of an enhancement $X(2075)$
near the threshold of the invariant mass spectrum of
${p}\bar{\Lambda}$ in $\jpsi\ar {p}{K^-}\bar{\Lambda}$ decays. The
mass, width, and product branching fraction of this enhancement are $M
= 2075 \pm 12~({\rm stat.}) \pm 5~({\rm syst.})$ MeV/$c^2$, $\Gamma =
90 \pm 35~({\rm stat.}) \pm 9~({\rm syst.})$ MeV/$c^2$, and $B(J/\psi
\ar K^- X)B(X \ar p \bar{\Lambda}+c.c.) = (5.9 \pm 1.4 \pm 2.0) \times
10^{-5}$~\cite{pkl}, respectively.  The study of the isospin conjugate
channel $\jpsi\ar n K^0_S \bar{\Lambda}$ is therefore important not
only in exploring new decay modes of $J/\psi$ but also in
understanding the $X(2075)$.


$J/\psi$ and $\psi(2S)\ar {n}{K^0_S}\bar{\Lambda}$ with ${K^0_S} \to
\pi^+\pi^-$ and $\bar{\Lambda} \to \bar p \pi^+$ (and $c.c.$) final
states contain four charged tracks and an undetected neutron or
anti-neutron.  We require the candidate events to have four charged
tracks with total charge zero. Secondary vertex fitting is used to
identify the $\pi^+ \pi^-$ from the $K^0_S$ and the $p \pi$ from the
$\Lambda$, and their masses are required to be consistent with those
of the parent particles.  To reject backgrounds from channels
without a $K^0_S$ or $\Lambda$, we require $L_{xy}(\Lambda)$, the
distance from the reconstructed $\Lambda$ vertex to the event origin,
to be larger than 5 mm and $L_{xy}(K^0_S) >5$ mm.  To suppress
background and improve the resolution, a one constraint (1C) kinematic
fit with a missing neutron is applied under the $J/\psi \to \bar p n
\pi^+\pi^-\pi^+$ hypothesis, and $\chi_{1C}^2 < 5$ is required. 

The invariant mass spectra of ${\Lambda K_S^0}$, ${n
K_S^0}$, and ${\bar {\Lambda} n (\Lambda \bar n)}$, as well as the
Dalitz plot for all selection requirements are shown in
Fig.~\ref{Nstar}.  In the $\Lambda K_S^0$ invariant mass spectrum, an
enhancement near $\Lambda K_S^0$ threshold is evident, as is found in
the $\Lambda K$ mass spectrum in $J/\psi \to p K^-
\bar{\Lambda}$~\cite{pkl2}.  The $X(2075)$ signal which was
seen in the $p\bar{\Lambda}$ invariant mass spectrum in $\jpsito p K^-
\bar{\Lambda}$ is not significant here. 
Taking into account the systematic error, the upper limit of the 
near-threshold enhancement $X(2075)$ in the $n\bar\Lambda$ 
threshold is
$B(\jpsi \ar {K^0_S} X(2075)) \cdot B( X(2075) \ar {n} \bar{\Lambda})
<4.9\times 10^{-5}$ (90\% C.L.).
Considering the isospin factor, the branching fraction upper 
limit for  
$B(\jpsi \ar {K^0_S} X) \cdot B( X \ar {n} \bar{\Lambda}+c.c.)$
is not inconsistent with that for 
$B(\jpsi \ar K X) \cdot B( X \ar {p} \bar{\Lambda}+c.c.)$~\cite{pkl}.

An $N^*$ state at around 1.9 GeV/c$^2$ in the $\Lambda K_S^0$ invariant
mass spectrum and $\Lambda^*$ states at around 1.5 and 1.7 GeV/c$^2$
in the $n K^0_S$ invariant mass spectrum are present. A larger data
sample and a partial wave analysis are needed to analyze these 
states.

We use the same criteria to
select $\psip\ar {n}{K^0_S}\bar{\Lambda}+c.c.$ events from the BESII sample of
14M $\psip$ events.
The 
branching ratios obtained are:
\begin{center}
$Br(\jpsi\ar n K^0_S\bar\Lambda+c.c)=(6.46\pm0.20\pm1.07)\times 10^{-4}$\\
$Br(\jpsi\ar n K^0_S\bar\Lambda)=(3.09\pm0.14\pm0.58)\times 10^{-4}$\\
$Br(\jpsi\ar \bar n K^0_S\Lambda)=(3.39\pm0.15\pm0.48)\times 10^{-4}$\\
$Br(\psip\ar n K^0_S\bar\Lambda+c.c)=(0.81\pm0.11\pm0.14)\times 10^{-4}$.
\end{center}
The ratio of the branching ratios
of $\psip$ and $\jpsi$ decaying to $n K^0_S\bar{\Lambda}+c.c$,
$Q_{h}$=(12.6$\pm$3.5)\%, obeys the "12\%" rule~\cite{rule}.
For more detail, see Ref.~\cite{nksL}.

\subsection{\boldmath $\psi(3770)$, $\psi(4040)$, $\psi(4160)$
and $\psi(4415)$ resonance parameters}

The total cross section for hadron production in $e^+e^-$ annihilation
is usually parameterized in terms of the ratio $R$, which is defined
as $R=\sigma(e^+e^- \rightarrow \mbox{hadrons})/
\sigma(e^+e^-\rightarrow \mu^+\mu^-)$, where the denominator is the
lowest-order QED cross section, $\sigma
(e^+e^-\rightarrow\mu^+\mu^-)=\sigma^0_{\mu \mu}=4\pi\alpha^2 / 3s$.
At the open flavor thresholds where resonance structures show up,
R measurements are used to determine resonance
parameters. For the high mass charmonium resonances, the $\psi(3770)$
was measured by MARK-I \cite{MarkI3770}, DELCO \cite{DELCO}, MARK-II
\cite{MarkII} and BES \cite{bes3770}\cite{besnondd}; the $\psi(4040)$
and $\psi(4160)$ were measured by DASP~\cite{DASP}; and the
$\psi(4415)$ was measured by DASP \cite{DASP} and MARK-I
\cite{MarkI4415}.

The most recent and precise $R$ measurements between 2-5 GeV were made
by BESII~\cite{besr99}. Experimentally, $R$ for both the continuum and
the wide resonance region is given by
\begin{equation}
R_{exp}=\frac{ N^{obs}_{had} - N_{bg}} { \sigma^0_{\mu\mu}  L
\epsilon_{trg} \epsilon_{had}(1+\delta_{obs})}, \label{rexp}
\end{equation}
where $N^{obs}_{had}$ is the number of observed hadronic events,
$N_{bg}$ is the number of the residual background events, $L$ is the
integrated luminosity, $(1+\delta_{obs})$ is the effective correction
factor of the initial state radiation (ISR)~\cite{CB}\cite{isrhu},
$\epsilon_{had}$ is the detection efficiency for hadronic events
determined by Monte Carlo simulation without bremsstrahlung being
simulated, and $\epsilon_{trg}$ is the trigger efficiency. 

\begin{table*}[htbp]
\caption{The resonance parameters of the high mass charmonia in this
work together with the values in PDG2004~\cite{pdg04},
PDG2006~\cite{pdg06} and K. Seth's evaluations~\cite{seth} based on
Crystal Ball and BES data.}
\vskip -0.1cm \hskip 1.5cm
\begin{center}
\footnotesize
\begin{tabular}{|c|c|c|c|c|c|} \hline \hline
        &         &$\psi(3770)$  &$\psi(4040)$&$\psi(4160)$&$\psi(4415)$\\ \hline
        &PDG2004  &3769.9$\pm$2.5&4040$\pm$10 &4159$\pm$20 & 4415$\pm$6 \\
        &PDG2006  &3771.1$\pm$2.4&4039$\pm$1.0&4153$\pm$3  & 4421$\pm$4 \\
$M$     &CB (Seth) &   -          &4037$\pm$2  &4151$\pm$4  & 4425$\pm$6 \\
(MeV/$c^2$)   &BES (Seth)&   -          &4040$\pm$1  &4155$\pm$5  & 4455$\pm$6 \\
        &BES (this work) &3771.4$\pm$1.8& 4038.5$\pm$4.6 & 4191.6$\pm$6.0 & 4415.2$\pm$7.5\\\hline\hline

              &PDG2004  &23.6$\pm$2.7&52$\pm$10    &78$\pm$20   &     43$\pm$15   \\
              &PDG2006  &23.0$\pm$2.7&80$\pm$10    &103$\pm$8    &     62$\pm$20   \\
$\Gamma_{tot}$&CB (Seth) &  -        &85$\pm$10     &107$\pm$10&119$\pm$16 \\
(MeV)         &BES (Seth)&  -        &89$\pm$6     &107$\pm$16&118$\pm$35\\
              &BES (this work)      &25.4$\pm$6.5&81.2$\pm$14.4&72.7$\pm$15.1&73.3$\pm$21.2\\\hline\hline

             &PDG2004  &0.26$\pm$0.04&0.75$\pm$0.15&0.77$\pm$0.23&0.47$\pm$0.10 \\
             &PDG2006  &0.24$\pm$0.03&0.86$\pm$0.08&0.83$\pm$0.07&   0.58$\pm$0.07 \\
$\Gamma_{ee}$&CB (Seth) &   -         &0.88$\pm$0.11&0.83$\pm$0.08&0.72$\pm$0.11\\
(keV)        &BES (Seth)&   -         &0.91$\pm$0.13&0.84$\pm$0.13&0.64$\pm$0.23\\
             &BES (this work)      &0.18$\pm$0.04&0.81$\pm$0.20&0.50$\pm$0.27&0.37$\pm$0.14\\\hline\hline
$\delta$ (degree)&BES (this
work)&0&133$\pm$68&301$\pm$61&246$\pm$86\\\hline\hline
\end{tabular}
\label{table}
\end{center}
\end{table*}

\begin{figure*}
\begin{center}
\includegraphics[width=8cm,height=6cm]{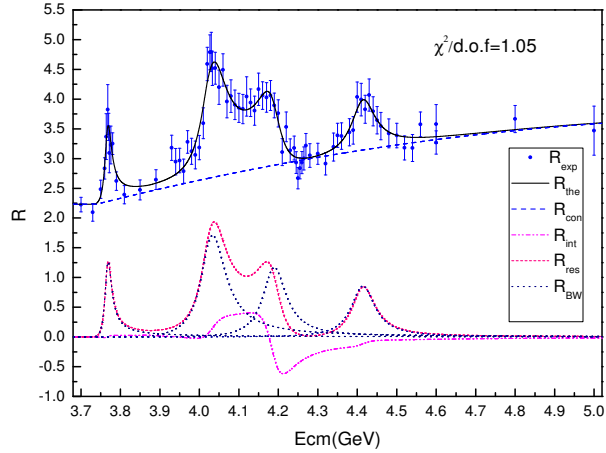}
\vspace{-5mm}\caption{The fit to the $R$ values in the high mass
charmonium region. The dots with error bars are the updated $R$
values. The solid curve shows the best fit, and the other curves
show the contributions from each resonance $R_{BW}$,  the
interference $R_{int}$, the summation of the four resonances
$R_{res}=R_{BW} + R_{int}$, and the continuum background $R_{con}$
respectively.} \label{resbes}
\end{center}
\end{figure*}

In the
previous analysis~\cite{besr99}, the determination of $R$ values was done using
PDG04~\cite{pdg04} resonance parameters for the high mass
resonances. Here, the analysis uses the data to determine the
resonance parameters.  The determination of $R$ values and resonance
parameters are intertwined; the factor $(1+\delta_{obs})$ in
Eq.~(\ref{rexp}) contains contributions from the resonances and
depends on the resonance parameters. Therefore, the procedure to
calculate $(1+\delta_{obs})$ requires a number of iterations before
stable results can be obtained.
We perform a global fit over the entire center-of-mass
energy region from 3.7 to 5.0 GeV covering the four resonances,
$\psi(3770)$, $\psi(4040)$, $\psi(4160)$ and $\psi(4415)$, and include
interference effects among the resonances. We also adopt
energy-dependent full widths, and introduce relative phases between
the resonances.


The resonant parameters of the high mass charmonia determined in
this work, together with those in PDG2004, PDG2006 and the results
given in Ref.~\cite{seth} are listed in Table~\ref{table}. The
updated $R$ values between 3.7 and 5.0 GeV and the fit curves are
illustrated in Fig.~\ref{resbes}.



It is worth noting that the change of the resonance parameters
affects the effective initial state radiative correction factors,
and thus affects the $R$ values.
In general the relative difference is within $3\%$, and for a few
energy points the maximum difference is about $6\%$. Our resonance
parameter results are in agreement with the previous experiments in
most cases, but large differences are observed in some of the
parameters, such as the mass of the $\psi(4160)$. This is mainly due
to the reconsideration of the radiative correction factors, and the
inclusion of interferences between the resonances.  This work is
preliminary; for more detail,
see Ref.~\cite{himassres}.


\section{BEPCII and BESIII}
 In 2003, the Chinese Government approved the upgrade of the BEPC to a
two-ring collider (BEPCII) with a design luminosity approximately 100
times higher than that of the BEPC.  This will allow unprecedented
physics opportunities in this energy region and contribute to
precision flavor physics.

\subsection{BEPCII}

BEPCII is a two-ring $e^+e^-$ collider that will run in the tau-charm
energy region ($E_{cm} = 2.0 - 4.2$ GeV) with a design luminosity of
$1\times 10^{33}$ cm$^{-2}$s$^{-1}$ at a beam energy of 1.89 GeV, an
improvement of a factor of 100 with respect to the BEPC. This is
accomplished by using multi-bunches and micro-beta. The upgrade
uses the existing tunnel. 

The 2024 meter long linac has been upgraded with new klystrons, a new
electron gun, and a new positron source to increase its energy and
beam current; it can accelerate electrons and positrons up to 1.89 GeV
with an positron injection rate of 50 mA/min. Its installation
was completed in the summer of 2005.

     
There are two storage rings with lengths of 237.5 meters.  The
collider has new super-conducting RF cavities, power supplies, and
control; super-conducting quadrupole magnets; beam pipes; magnets and
power supplies; kickers; beam instrumentation; vacuum systems; and
control.  The old dipoles are modified and used in the outer ring.
Electrons and positrons will collide at the interaction point with a
horizontal crossing angle of 11 mrad and bunch spacing of 8 ns.  Each
ring has 93 bunches with a beam current of 9170 mA.  The machine is
already providing a high flux of synchrotron radiation at a beam energy of
2.5 GeV.  

\subsection{BESIII}
\begin{figure}  \centering
   \includegraphics*[width=0.45\textwidth]{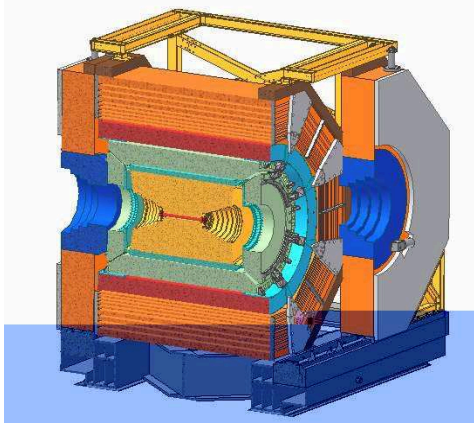}  
  \caption{\label{schematic}Schematic view of the BESIII detector.
    }
 \end{figure}

The BESIII detector consists of a berylium beam pipe, a helium-based
small-celled drift chamber, Time-Of-Flight counters for particle
identification, a CsI(Tl) crystal calorimeter, a super-conducting
solenoidal magnet with a field of 1 Tesla, and a muon identifier using
the magnet yoke interleaved with Resistive Plate Counters
(RPC). Fig.~\ref{schematic} shows the schematic view of the BESIII
detector, including both the barrel and endcap portions.
    
\subsection{Physics in the tau-charm energy region}

The tau-charm energy region makes available a wide variety of interesting
physics.  Data can be taken at the $J/\psi$, $\psi(2S)$, and
$\psi(3770)$, at $\tau$ threshold, and at an energy to allow production
of $D_s$ pairs, as well as for an R-scan.

\begin{table*}[htb]
\caption{\label{3.1-1}
 Number of events expected for one year of running.}
\begin{center}
\begin{tabular}[!h]{| l | c | c | c | c| }
\hline
Physics & Center-of-mass  &  Peak          & Physics      & Number of \\ 
channel & Energy          & Luminosity     & cross    & Events per \\ 
        & (GeV)  & ($10^{33}$ cm$^{-2}$ s$^{-1}$) & section (nb) & year\\\hline
$J/\psi$   &   3.097  &  0.6  &  $\sim 3400$ & $10\times 10^9$ \\
$\tau$     &   3.67   &  1.0  &  $\sim2.4$   & $12 \times 10^6$ \\
$\psi(2S)$ &   3.686  &  1.0  &  $\sim640$   & $3.0 \times 10^9$ \\
$D$        &   3.770  &  1.0  &  $\sim5$     & $25 \times 10^6 $\\
$D_s$      &   4.030  &  0.6  &  $\sim0.32$  & $1.0 \times 10^6$ \\
$D_s$      &   4.140  &  0.6  &  $\sim0.67$  & $2.0 \times 10^6$ \\
\hline \end{tabular}
\end{center}
\end{table*}

BEPCII and BESIII are in the final stage of assembly, and 
commissioning will begin in summer 2008.  The design luminosity of BESIII is $1
\times 10^{33}$ cm$^{-2}$.  Clearly BESIII with higher luminosity will
contribute greatly to precision flavor physics: $V_{cd}$ and $V_{cs}$
will be measured with a statistical accuracy of 1.6\%. $D^0 D^0$
mixing will be studied and CP violation will be searched
for. Table~\ref{3.1-1} gives the numbers of events expected during one
year of running at various energies.  Huge $J/\psi$ and $\psi(2S)$
samples will be obtained.  The $\eta_C$, $\chi_{CJ}$, and $h_C$ can be
studied with high statistics, and the $\rho \pi$ puzzle will be
studied with better accuracy. For more detail, see
Refs~\cite{fahbesiii,weiguobesiii}.




\section{References}





\end{document}